\documentclass[aps,prl,twocolumn,superscriptaddress,nofootinbib,notitlepage,longbibliography]{revtex4-2}
\usepackage{amsmath,amssymb,amsfonts,bm,color,graphicx,tabularx}

\usepackage[etex=true,export]{adjustbox}
\usepackage{makecell}
\usepackage{multirow}
\usepackage{amsmath}
\usepackage{verbatim}
\usepackage{longtable}
\usepackage[colorlinks=true,citecolor=cyan]{hyperref}
\hypersetup{colorlinks=true,citecolor=cyan,linkcolor=blue,urlcolor=magenta}
\usepackage{tabularx,multirow,array,diagbox}
\usepackage{adjustbox}
\usepackage{supertabular}

\newcommand{\beq}{\begin{equation}}
\newcommand{\eeq}{\end{equation}}

\begin{document}
\title{Chiral Majorana Charge Kondo Criticality}

\author{Xiao Xiao}
\affiliation{Department of Physics, Northeastern University, Boston, Massachusetts 02115, USA}
\affiliation{Quantum Materials and Sensing Institute, Northeastern University, Burlington, Massachusetts 01803, USA}

\begin{abstract}
Unambiguous identification of chiral Majorana edge modes remains challenging because the transport signatures proposed for them can also arise from conventional mechanisms. Here we identify the chiral Majorana charge Kondo (CMCK) critical point as a more discriminating many-body route to detect chiral Majorana edge modes. This criticality arises in a charged island hosting a single chiral Majorana edge mode, where Majorana self-conjugacy locks normal and anomalous tunneling amplitudes, so that each physical lead provides two Kondo screening channels without fine tuning. Ordinary subgap states generically break this locking and drive the system toward conventional charge Kondo (CK) behavior. In the simplest two-terminal geometry, CMCK and conventional CK share the critical conductance $G^*=e^2/2h$, but their leading conductance corrections scale as $T^{1/3}$ and $T$, respectively. Their residual entropies also differ, as do the crossover exponents under gate detuning and contact asymmetry. For $N$ leads, CMCK conductance corrections scale as $T^{1/(N+1)}$, and for $N\geq3$ even the critical conductance differs from that of conventional CK with the same number of leads. These universal signatures are testable using established charge-Kondo measurement techniques.
\end{abstract}

\maketitle

\emph{Introduction.} Majorana fermions, being their own antiparticles, lie at the intersection of topological quantum matter and quantum information science~\cite{Kitaev2001,Nayak2008,Alicea2012}. Chiral Majorana modes, predicted at the edges of two-dimensional topological superconductors and non-Abelian Kitaev spin liquids~\cite{Read2000,Qi2010,Kitaev2006}, offer an alternative route toward topological quantum computation~\cite{Lian2018,Beenakker2019}. Their identification, however, remains challenging because conventional mechanisms can mimic proposed signatures: zero-bias conductance peaks~\cite{Law2009} can arise from trivial Andreev bound states~\cite{Kells2012,Liu2017,Prada2020}, a reported half-quantized conductance plateau attributed to chiral Majorana modes was retracted after alternative interpretations and follow-up experiments~\cite{He2017,Ji2018,Sau2018,Kayyalha2020,Thorp2022}, and the half-quantized thermal Hall conductance in $\alpha$-RuCl$_3$ remains controversial~\cite{Kasahara2018,Yokoi2021,Yamashita2020,Czajka2021,Lefrancois2022}. Probes based on universal many-body properties, beyond single-particle transport or interferometry~\cite{Law2009,FuKane2009,Akhmerov2009,WangJ2015,Aasen2020}, are therefore highly desirable.

The charge Kondo (CK) effect provides such a probe~\cite{Matveev1991,Furusaki1995}: in a metallic island coupled to leads through quantum point contacts, two degenerate charge states act as a pseudospin and the leads as Kondo channels. Two- and three-channel CK critical points have been realized with universal conductances and exponents in quantitative agreement with conformal field theory (CFT)~\cite{Iftikhar2015,Mitchell2016,BaoZhang2017,Iftikhar2018}, and fractional residual entropies have recently been extracted~\cite{HanC2022,Piquard2026}.

\begin{figure}[h]
  \centering
  \includegraphics[width=\columnwidth]{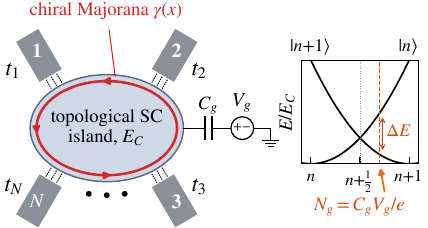}
  \caption{Schematic setup: a charged topological superconducting island with charging energy $E_C$ hosts a chiral Majorana edge mode $\gamma(x)$ and is tunnel-coupled to $N$ leads with coupling strength $t_j$. A gate voltage $V_g$ applied through the capacitance $C_g$ sets $N_g=C_g V_g/e$. Inset: charging energies of the $|n\rangle$ and $|n+1\rangle$ states. At charge degeneracy, $N_g = n+1/2$, the two states form a pseudospin, and a detuning $\Delta E$ acts as an external Zeeman field.}
  \label{fig1}
\end{figure}

In this Letter, we show that replacing the metallic island of the conventional charge Kondo setup with a charged island supporting a chiral Majorana edge mode gives rise to a distinct quantum critical point (Fig.~\ref{fig1}), which we term the chiral Majorana charge Kondo (CMCK) critical point. The self-conjugacy of the chiral Majorana mode locks normal and anomalous tunneling amplitudes together, so that each physical lead provides two Kondo screening channels without fine tuning, as in channel-isotropic Kondo models proposed for quantum dots coupled to Read-Rezayi states~\cite{Fiete2008,Sevier2011}. This enforced channel doubling yields an overscreened $2N$-channel Kondo sector for $N$ leads, with each lead acting as two perfectly matched channels: two leads already realize four-channel criticality, beyond the two- and three-channel critical points demonstrated so far~\cite{Iftikhar2015,Iftikhar2018} and complementing other recent routes to higher-channel Kondo physics~\cite{Gaines2026}.

Kondo effects in floating Majorana islands~\cite{Fu2010} have so far relied on localized zero modes, which form the impurity either as an $SO(M)$ spin in a Coulomb valley~\cite{Beri2012,Altland2013,Li2023} or near charge degeneracy~\cite{Herviou2016,Michaeli2017,Lutchyn2017,Vayrynen2020}. In the CMCK setup, by contrast, the impurity is the charge pseudospin alone, while the Majorana is a gapless propagating mode of the bath. Coulomb-blockaded chiral Majorana islands have been studied in the weak-tunneling, high-bias regime~\cite{Shapiro2023}, whereas the criticality identified here emerges at low energies, with universal transport properties set only by the number of leads. This universality turns transport into a diagnostic of the Majorana edge itself, a strategy with precedent in the two-channel Kondo effect predicted for Majorana Kramers pairs~\cite{BaoZhang2017PRL}, whose experimental realization is still under exploration. The CMCK critical point brings this strategy to chiral Majorana edge modes in quantum anomalous Hall (QAH)-superconductor devices, a platform already under active investigation~\cite{Uday2024} and in need of discriminating experimental tests.

The resulting criticality is experimentally distinguishable even in the simplest two-terminal ($N=2$) geometry. Although CMCK and conventional CK share the critical conductance $G^*=e^2/2h$, they approach it differently, with leading corrections scaling as $T^{1/3}$ rather than $T$, and the residual impurity entropy is $S_{\rm imp}=\frac{1}{2}\ln 3$ rather than $\frac{1}{2}\ln 2$. Their responses to perturbations provide further tests. Contact asymmetry produces a smaller crossover scale, $T^*_{\rm LR}\propto|\delta|^3$ instead of $|\delta|^2$, whereas charge detuning yields $T^*_\Delta\propto|\Delta E|^{3/2}$ instead of $|\Delta E|^2$. These exponents coincide with those of conventional four-lead CK, both having four screening channels, but the lead-resolved conductance matrices differ, because the symmetric-mode projection acts on two physical leads in CMCK rather than four. Ordinary subgap states break the Majorana-enforced tunneling constraint and drive the system toward conventional CK. Independently, tuning the same island to its QAH phase provides an in situ conventional-CK baseline for comparison. We establish these results through an explicit Sugawara construction and the Affleck-Ludwig boundary CFT framework~\cite{AffleckLudwig1991NPB,AffleckLudwig1991PRL,AffleckLudwig1993,vonDelft1995}. The construction extends to generic $N$, for which even the critical conductance differs from that of conventional CK with the same number of leads when $N\geq3$, providing an additional test in three-terminal circuits~\cite{Iftikhar2018,Paris2026}.

\emph{Setup and model.} We focus on the setup in Fig.~\ref{fig1}, where the island is a quantum anomalous Hall (QAH) insulator-superconductor heterostructure~\cite{Qi2010,WangJ2015}, though a gate-defined charging region coupled to a non-Abelian Kitaev spin liquid is a possible
alternative \cite{Kitaev2006,Halasz2024}. Tuning the magnetization drives a transition from the QAH phase, which hosts a chiral Dirac edge mode, to the chiral topological superconducting (CTS) phase, in which the Dirac mode splits into two chiral Majorana modes, one merging into the bulk and the other remaining at the island boundary. Near each well-separated point contact $0_j$, the leads and the island edge are described by chiral fermions $\psi_{j,l}$ and $\psi_{j,f}$ with a common velocity $v$, $H_0 = \sum_{j,\alpha} \int dx\, \psi^\dagger_{j,\alpha}(-iv\partial_x)\psi_{j,\alpha}$, and the island is charged by $H_C = E_C(\hat{n} - N_g)^2$ with $N_g = C_g V_g/e$. For $E_C \gg k_B T$ and $N_g \simeq n + \frac{1}{2}$, only $|n\rangle$ and $|n+1\rangle$ are retained, defining the pseudospin $S_- = |n\rangle\langle n+1| = S_+^\dagger$, and the detuning $\Delta E = E_C(2n+1-2N_g)$ enters as $\Delta E\, S^z$. Writing $\psi_{j,f} = (\gamma_{j;1} + i\gamma_{j;2})/\sqrt{2}$, the tunneling reads
\beq \label{Eq1}
H_T = \sum_j \left[ \psi_{j,l}^\dagger(0_j)\, \frac{t_{j;1}\gamma_{j;1}(0_j) + i t_{j;2}\gamma_{j;2}(0_j)}{\sqrt{2}}\, S_- + {\rm h.c.} \right].
\eeq
For a grounded island, Eq. (1) describes resonant Andreev reflection through the chiral Majorana mode \cite{Law2009}. Introducing the spin and isospin densities of the lead-edge doublet $(\psi_{j,l},\psi_{j,f})$ at $0_j$,
$\sigma_j^+ = \psi_{j,l}^\dagger\psi_{j,f}$, $\sigma_j^z = \frac{1}{2}(\psi_{j,l}^\dagger\psi_{j,l} - \psi_{j,f}^\dagger\psi_{j,f})$ and
$\tau_j^+ = \psi_{j,l}^\dagger\psi_{j,f}^\dagger$, $\tau_j^z = \frac{1}{2}(\psi_{j,l}^\dagger\psi_{j,l} + \psi_{j,f}^\dagger\psi_{j,f} - 1)$, Eq.~(\ref{Eq1}) becomes
\beq \label{Eq2}
H_T = \sum_j \left[ t_{j;s}\left(\sigma_j^+ S_- + {\rm h.c.}\right) + t_{j;a}\left(\tau_j^+ S_- + {\rm h.c.}\right) \right],
\eeq
with $t_{j;s/a} = (t_{j;1} \pm t_{j;2})/2$. In the QAH phase, $t_{j;1} = t_{j;2}$, so $t_{j;a} = 0$ and only the spin $\boldsymbol{\sigma}_j$ couples to $\mathbf{S}$, recovering the conventional CK model~\cite{Matveev1991,Furusaki1995}. In the CTS phase, $\gamma_{j;2}$ merges into the bulk, so $t_{j;2} = 0$ and $t_{j;s} = t_{j;a}$: spin and isospin couple to $\mathbf{S}$ with equal strength. 

\emph{The CMCK critical point.} To identify the fixed points of Eq.~(\ref{Eq2}), we perform a perturbative renormalization group (RG) analysis at weak tunneling. Besides renormalizing the tunneling amplitudes, the RG generates longitudinal couplings between the pseudospin and the quasi-particles on both leads and edge,
\beq \label{Eq5}
O_j^z = \frac{1}{2}\left[ t^z_{j;l}\,\psi^\dagger_{j,l}\psi_{j,l} - t^z_{j;f}\,\psi^\dagger_{j,f}\psi_{j,f} \right](0_j)\, S^z .
\eeq
For identical contacts, as realized in charge-Kondo experiments~\cite{Iftikhar2015,Iftikhar2018}, the two-loop flow equations for the dimensionless couplings $g_{j;s/a} = t_{j;s/a}\rho_0$ and $g^z_{j;l/f} = t^z_{j;l/f}\rho_0$, with $\rho_0$ the density of states, read (see the Supplemental Material (SM)~\cite{supporting})
\begin{align} \label{Eq6}
\frac{dg^z_l}{d\ell} &= \left(g_s^2+g_a^2\right)\left(1-\frac{N}{2}\,g^z_l\right), \nonumber\\
\frac{dg^z_f}{d\ell} &= \left(g_s^2-g_a^2\right) - \frac{N}{2}\,g^z_f\left(g_s^2+g_a^2\right), \nonumber\\
\frac{dg_s}{d\ell} &= g_s\left[\frac{g^z_l+g^z_f}{2} - \frac{N}{4}\,\Gamma\right], \nonumber\\
\frac{dg_a}{d\ell} &= g_a\left[\frac{g^z_l-g^z_f}{2} - \frac{N}{4}\,\Gamma\right],
\end{align}
with $\Gamma = \frac{1}{2}[(g^z_l)^2+(g^z_f)^2] + g_s^2 + g_a^2$. Eq.~(\ref{Eq6}) reveals two intermediate-coupling fixed points, $(g^z_l, g^z_f, g_s, g_a) = (\frac{2}{N}, \frac{2}{N}, \pm\frac{2}{N}, 0)$ and $(\frac{2}{N}, 0, \pm\frac{1}{N}, \pm\frac{1}{N})$. The first is the conventional $N$-channel CK fixed point, and the second, at which the edge density decouples and normal and anomalous tunneling become equal, is the CMCK fixed point. Since both occur at couplings of order $1/N$, the two-loop RG, which is controlled only for $N \gg 1$, is to identify the fixed-point structure in the parameter space.

\begin{figure*}[t]
  \centering
  \includegraphics[width=1.8\columnwidth]{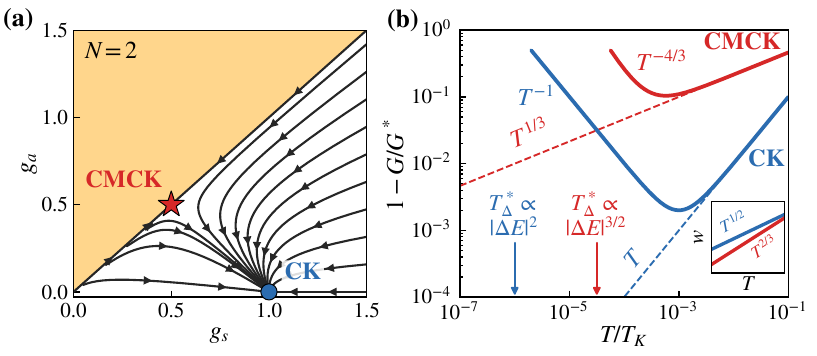}
  \caption{(a) Two-loop RG flow in the $(g_s,g_a)$ plane for $N=2$: the Majorana line $g_s=g_a$ flows into the CMCK fixed point (star), while all other flows end at the conventional CK fixed point (circle). (b) Deviation from the fixed-point conductance at detuning $\Delta E/T_K=10^{-3}$. Dashed lines: $\Delta E=0$, with $1-G/G^* \propto T^{1/3}$ for CMCK fixed point and $\propto T$ for CK fixed point. Below $T_\Delta^* \propto |\Delta E|^{1/y_\Delta}$ (arrows), the system leaves the fixed point. Inset: Coulomb peak width, $w \sim T^{y_\Delta}$ with $y_\Delta=2/3$ for CMCK and $y_\Delta=1/2$ for CK.}
  \label{fig2}
\end{figure*}

The Majorana manifold with $g^z_f = 0$ and $g_s = g_a$ is invariant under Eq.~(\ref{Eq6}), and the CMCK fixed point is attractive within it (see Fig.~\ref{fig2}(a)). In the ideal CTS phase, $t_{j;2} = 0$ is enforced by the Majorana structure of the edge, so no fine tuning is required. Perturbations violating this constraint, such as residual coupling to $\gamma_{j;2}$ or to ordinary subgap states, constitute a single relevant direction with RG eigenvalue $y_M = 1/N$, which generates a crossover scale $T_M^* \sim T_K |\delta_M|^N$, where $\delta_M$ is the deviation from the manifold at $T_K$ (see SM~\cite{supporting}). For an ideal Majorana edge, $T_M^* = 0$ and the CMCK fixed point controls the zero-temperature limit. Otherwise, CMCK scaling persists for $T_M^* \ll T \ll T_K$ before the flow crosses over to the conventional fixed point.

To determine universal features, we focus on the Majorana manifold, where the interaction takes the isotropic Kondo form:
\beq \label{Eq7}
H^* = H_0 + t \sum_j \mathbf{J}_j(0_j)\cdot\mathbf{S} + \Delta E\, S^z, \qquad \mathbf{J}_j = \boldsymbol{\sigma}_j + \boldsymbol{\tau}_j,
\eeq
where a term $\propto t\,S^z$ generated by the RG has been absorbed into $\Delta E$.

\emph{The current algebra for $N=2$.} We first focus on the charge-degeneracy point, $\Delta E = 0$, and discuss the effect of detuning later. To identify the critical point, we decompose each contact into Majorana fermions, $\psi_{j,l} = (\gamma_{j,1} - i\gamma_{j,2})/\sqrt{2}$ and $\psi_{j,f} = -(\gamma_{j,3} + i\gamma_{j,4})/\sqrt{2}$, i.e., $\gamma_{j,3,4} = -\gamma_{j;1,2}$, so that $H_0 = \frac{v}{2}\sum_{j,a}\int dx\,\gamma_{j,a}(-i\partial_x)\gamma_{j,a}$, and the currents $\mathbf{J}_j = \boldsymbol{\sigma}_j + \boldsymbol{\tau}_j$ that couple to the pseudospin in Eq.~(\ref{Eq7}) become
\beq \label{Eq8}
J_j^a = -\frac{i}{2}\,\epsilon^{abc}\gamma_{j,b}\gamma_{j,c}, \qquad a,b,c \in \{1,2,3\}.
\eeq
Only the triplet $\gamma_{j,1,2,3}$ couples to the pseudospin, while $\gamma_{j,4}$ decouples, and the impurity coupling of Eq.~(\ref{Eq7}) reads $t\,\mathbf{J}(0)\cdot\mathbf{S}$ with $\mathbf{J} = \mathbf{J}_1 + \mathbf{J}_2$. Each $\mathbf{J}_j$ generates an $SU(2)_2$ algebra, and the total current $\mathbf{J}$ generates $SU(2)_4$,
\beq \label{Eq9}
[J^a_n, J^b_m] = i\epsilon^{abc}J^c_{n+m} + 2n\,\delta^{ab}\delta_{n+m,0}.
\eeq
The Sugawara term $\mathbf{J}\cdot\mathbf{J}$ alone does not reproduce the kinetic energy of the six coupled Majoranas, because it contains four-Majorana terms mixing the two contacts. These are canceled by the $U(1)$ current
\beq \label{Eq10}
\mathcal{J} = i\sum_{a=1}^{3}\gamma_{1,a}\gamma_{2,a}, \qquad [\mathcal{J}_n, \mathcal{J}_m] = 3n\,\delta_{n+m,0},
\eeq
which generates rotations between the Majorana triplets of the two contacts and commutes with $\mathbf{J}$. Omitting the decoupled Majoranas $\gamma_{j,4}$, we then find
\beq \label{Eq11}
H^* = \frac{\pi v}{3}\int dx\left[{:}\mathcal{J}\mathcal{J}{:} + {:}\mathbf{J}\cdot\mathbf{J}{:} \right] + t \mathbf{J}(0)\cdot\mathbf{S}.
\eeq
The six coupled Majoranas thus decompose into $U(1)_3 \times SU(2)_4$, with central charges $3 = 1 + 2$. The impurity couples only to the modes of the $SU(2)_4$ current, while the $U(1)_3$ sector, which commutes with $\mathbf{J}$, is a spectator. The problem therefore reduces to a spin-$\frac{1}{2}$ impurity coupled to an $SU(2)_4$ current. Since the level exceeds twice the impurity spin, $k = 4 > 2S = 1$, the impurity is overscreened.

\emph{Absorbing the impurity.} Writing Eq.~(\ref{Eq11}) in terms of the current modes, the $SU(2)_4$ sector and the impurity read
\beq \label{EqHmodes}
H^*_{SU(2)} = \frac{2\pi v}{L}\sum_n\left[\frac{1}{6}\,{:}\mathbf{J}_{-n}\cdot\mathbf{J}_n{:} + \frac{t}{2\pi v}\,\mathbf{J}_n\cdot\mathbf{S}\right],
\eeq
while the $U(1)_3$ sector decouples. Following Affleck and Ludwig~\cite{AffleckLudwig1991NPB,AffleckLudwig1991PRL},
and their generalizations to nontrivial current-algebra decompositions~\cite{vonDelft1998,Walter2020}, we absorb the impurity into the shifted currents $\tilde{\mathbf{J}}_n = \mathbf{J}_n + \alpha \mathbf{S}$. Since $[S^a, S^b] = i\epsilon^{abc}S^c$, the $\tilde{\mathbf{J}}_n$ obey the same $SU(2)_4$ algebra, Eq.~(\ref{Eq9}), only if $\alpha^2 = \alpha$, which fixes $\alpha = 1$. At the special coupling
\beq \label{Eqtstar}
\frac{t^*}{2\pi v} = \frac{2}{k+2} = \frac{1}{3},
\eeq
Eq.~(\ref{EqHmodes}) becomes $\frac{2\pi v}{L}\sum_n\frac{1}{6}{:}\tilde{\mathbf{J}}_{-n}\cdot\tilde{\mathbf{J}}_n{:}$ up to a constant. The impurity thus disappears into a free Sugawara Hamiltonian of the shifted currents, which differ from $\mathbf{J}_n$ only in that their $SU(2)$ quantum numbers include the spin $\frac{1}{2}$ of the island. Expanding $t^*/2\pi v = \frac{2}{k}(1 - \frac{2}{k} + \cdots)$ with $k = 2N$, the leading term reproduces the weak-coupling CMCK fixed point of Eq.~(\ref{Eq6}), which connects the RG analysis to the strong-coupling description.

\emph{Critical properties.} After absorbing $\mathbf{S}$ into the $SU(2)_4$ current, the new conformal towers are obtained by fusing the $SU(2)_4$ sector with the $j = \frac{1}{2}$ impurity, leaving the $U(1)_3$ sector untouched (see SM~\cite{supporting} for the conformal towers and gluing conditions). The residual impurity entropy then follows from the modular $S$ matrix,
\beq \label{EqSimp}
S_{\rm imp} = \ln\frac{S_{1/2,0}}{S_{0,0}} = \ln\frac{\sin(\pi/3)}{\sin(\pi/6)} = \frac{1}{2}\ln 3,
\eeq
which is distinct from $S_{\rm imp} = \frac{1}{2}\ln 2$ of the conventional $N=2$ CK critical point.

The conductance matrix is determined by the boundary amplitudes $R_{ij}$, which correlate the outgoing density in lead $i$ with the incoming density in lead $j$, through $G_{ij} = \frac{e^2}{h}(\delta_{ij} - R_{ij})$ \cite{WongAffleck1994,SelaAffleck2009}. It is convenient to work in the symmetric and antisymmetric combinations $j_\pm = (j_1 \pm j_2)/\sqrt{2}$, in which the boundary amplitudes are diagonal. The symmetric current $j_+ = J^z/\sqrt{2}$ belongs to the vacuum conformal tower and is therefore transmitted with $R_+ = 1$. By contrast, the antisymmetric current $j_-$ carries $U(1)_3$ charge $\pm 2$ and lies in the $SU(2)_4$ $j = 1$ tower, for which $R_- = (S_{1,1/2}/S_{1,0})/(S_{0,1/2}/S_{0,0}) = 0$. Transforming back to the physical lead basis gives:
\beq \label{N=2_conductance}
G = \frac{e^2}{2h}\begin{pmatrix} 1 & -1 \\ -1 & 1 \end{pmatrix}.
\eeq
The two-terminal conductance at the CMCK critical point is thus $G^* = e^2/2h$. It coincides with that of the conventional $N=2$ CK critical point, but here it arises from the vanishing of the modular S matrix element $S_{1,1/2}$ for $SU(2)_4$.

The leading corrections to the CMCK fixed point are governed by irrelevant boundary operators contained in the fusion product $\frac{1}{2}\times\frac{1}{2} = 0\oplus 1$. Imposing the pseudospin, $U(1)_3$ charge conservation, and lead-exchange symmetry, one can identify the unique singlet $\tilde{\mathbf{J}}_{-1}\cdot\boldsymbol{\phi}$, built from the $j = 1$ primary $\boldsymbol{\phi}$ of dimension $\frac{1}{3}$, with $\Delta_{\rm irr} = \frac{4}{3}$. Its three-point coefficient with the antisymmetric current operators entering the conductance is proportional to a quantum $6j$ symbol of $SU(2)_4$ and does not vanish (see SM~\cite{supporting}), so the leading conductance correction appears already at first order in the irrelevant coupling. Scale invariance then gives
\beq \label{EqdG}
G(T, V) = G^* + A\left(\frac{T}{T_K}\right)^{1/3}\mathcal{F}\!\left(\frac{e|V|}{T}\right),
\eeq
with a universal scaling function $\mathcal{F}$ \cite{vonDelft1995} that approaches a constant for $e|V| \ll T$ and grows as $\mathcal{F}(x) \propto x^{1/3}$ for $x \gg 1$, so that the conductance at $T = 0$ becomes independent of temperature, $G(0, V) = G^* + B\,(e|V|/T_K)^{1/3}$, with nonuniversal constants $A$ and $B$. These contrast to the conventional $N=2$ CK critical point, where
the first-order contribution is forbidden by the $SU(2)_2$ fusion rule
$1\times1=0$, giving rise to corrections linear in $T$ and
$eV$~\cite{Iftikhar2015,Mitchell2016,BaoZhang2017,supporting,Li2026}.

Two perturbations are relevant. The charge detuning $\Delta E\,S^z$ maps onto $\phi^z$, with RG eigenvalue $y_\Delta = \frac{2}{3}$, which gives the crossover scale
\beq \label{EqTDelta}
T^*_\Delta = T_K\left(\frac{|\Delta E|}{T_K}\right)^{3/2}.
\eeq
Its first-order correction to $G$ vanishes by pseudospin symmetry, so that $\delta G \propto (\Delta E/T_K)^2(T_K/T)^{4/3}$ for $T \gg T^*_\Delta$, and the width of the Coulomb peak scales as $w \propto T^{2/3}$, compared with $T^{1/2}$ at the $N=2$ CK critical point [Fig.~\ref{fig2}(b)]. 

A difference between the couplings, $\delta = (t_1 - t_2)/(t_1 + t_2)$, enters through $(\mathbf{J}_1 - \mathbf{J}_2)\cdot\mathbf{S}$, which carries $U(1)_3$ charge $\pm 2$ and, after fusion, maps to a pseudospin singlet of dimension $\frac{2}{3}$. This yields $T^*_{\rm LR} = T_K\,|\delta|^3$ compared with $T_K|\delta|^2$ at the $N=2$ CK critical point, so the CMCK fixed point is more robust against contact asymmetry. The corresponding correction, $\delta G \propto \delta^2(T_K/T)^{2/3}$, appears only at second order, since a first-order perturbation cannot conserve the $U(1)_3$ charge.

\begin{table}[t]
\caption{\label{tab1}Universal properties of the CMCK and the conventional CK critical points~\cite{Iftikhar2015,Mitchell2016,BaoZhang2017} for $N = 2$.}
\begin{ruledtabular}
\begin{tabular}{lcc}
 & CMCK & CK \\
\hline
Conductance $G^*$ & $e^2/2h$ & $e^2/2h$ \\
Residual entropy $S_{\rm imp}$ & $\frac{1}{2}\ln 3$ & $\frac{1}{2}\ln 2$ \\
Correction $G - G^* \propto T^{x}$ & $T^{1/3}$ & $T$ \\
Peak width $w \propto T^{y_\Delta}$ & $T^{2/3}$ & $T^{1/2}$ \\
Detuning scale $T^*_\Delta$ & $|\Delta E|^{3/2}$ & $|\Delta E|^{2}$ \\
Asymmetry scale $T^*_{\rm LR}$ & $|\delta|^{3}$ & $|\delta|^{2}$ \\
\end{tabular}
\end{ruledtabular}
\end{table}

\emph{Generic $N$ leads.} The $N=2$ construction generalizes directly. Each contact contributes its triplet $\gamma_{j,1,2,3}$, so the $3N$ coupled Majoranas realize $SO(3N)_1$, whose central charge is $3N/2$. The intra-contact currents
$\mathbf{J}=\sum_j\mathbf{J}_j$ rotate the flavor index and generate
$SU(2)_{2N}$, while the inter-contact currents
$\mathcal{J}^{ij}=i\sum_a\gamma_{i,a}\gamma_{j,a}$ rotate the contact
index and generate $SO(N)_3$, generalizing $U(1)_3\simeq SO(2)_3$. The
central charges of the two, $3N/(N+1)$ and $3N(N-1)/[2(N+1)]$, add up to $3N/2$ (see SM~\cite{supporting} for the explicit Sugawara construction). The impurity again couples only to $SU(2)_{2N}$ and is absorbed at $t^*/2\pi v=1/(N+1)$.

Fusion with $j=\frac12$ gives $S_{\rm imp}=\ln[2\cos\frac{\pi}{2(N+1)}]$.
The symmetric current $\sum_i j_i\propto J^z$ lies in the vacuum tower,
$R_{\rm sym}=1$, while the $N-1$ orthogonal combinations are $SU(2)$
vectors with $R_{\rm rel}=\cos\frac{3\pi}{2(N+1)}/\cos\frac{\pi}{2(N+1)}$.
Transforming back to the lead basis gives
\begin{equation}
G^*_{i,j}=\frac{4e^2}{h}\sin^2\!\left[\frac{\pi}{2(N+1)}\right]
\left(\delta_{i,j}-\frac{1}{N}\right),
\end{equation}
which reduces to Eq.~(\ref{N=2_conductance}) for $N=2$.

The $j=1$ primary $\phi$ now has dimension $1/(N+1)$, so the leading
irrelevant operator $\tilde{\mathbf{J}}_{-1}\cdot\phi$ yields corrections
$\propto T^{1/(N+1)}$, and detuning, with $y_\Delta=N/(N+1)$, gives
$T^*_\Delta=T_K(|\Delta E|/T_K)^{(N+1)/N}$ and $w\propto T^{N/(N+1)}$.
Contact asymmetry, $(\mathbf{J}_i-\mathbf{J}_j)\cdot\mathbf{S}$,
transforms as a symmetric traceless tensor of $SO(N)_3$, generalizing the
$U(1)_3$ charge $\pm2$, and becomes after fusion a pseudospin singlet of
dimension $N/(N+1)$, so that $T^*_{\rm LR}=T_K|\delta|^{N+1}$.

\emph{Experimental realization.} We focus on $N=2$~\cite{Iftikhar2015}, while the three-contact geometry of Ref.~\cite{Iftikhar2018} can further test the $N\geq3$ critical conductance. Unlike an ordinary superconducting island, where odd-charge states cost the gap, the gapless Majorana edge supplies zero-energy fermionic states, so that $|n\rangle$ and $|n+1\rangle$ can be tuned into degeneracy as for a metallic island. Observing CMCK scaling requires $\max(k_BT^*,\delta_{\rm edge}) \ll k_BT \ll k_BT_K \ll \min(E_C,\Delta_{\rm topo})$, where $T^*=\max(T^*_\Delta,T^*_{LR},T^*_M)$ collects the relevant crossover scales, $\Delta_{\rm topo}$ is the induced topological gap, and $\delta_{\rm edge}\simeq 2\pi\hbar v/L$ is the edge level spacing for island perimeter $L$. The lower bound makes the edge a continuum on the thermal scale and, since the contacts are separated by $d\sim L/2$, keeps them independent. Finite-bias scaling additionally requires $e|V| \ll k_B T_K$. Two features are favorable: $T^*_M$ vanishes for an ideal Majorana edge, and $T^*_{LR}\propto|\delta|^3$ relaxes the contact-symmetry requirement relative to conventional two-terminal CK.

We further estimate the individual scales based on existing experiments. Metallic charge-Kondo devices reach $E_C\simeq k_B\times 290$~mK at electronic temperatures $T\simeq 11.5$~mK, with $T_K$ approaching $E_C/k_B$ at high contact transmission~\cite{Iftikhar2015}. Superconducting correlations induced in the chiral edge of QAH films have been detected at 17-25~mK, while the longitudinal resistance becomes finite above approximately 50~mK, signaling bulk contributions~\cite{Uday2024}. Representative design scales are thus $E_C/k_B\sim 0.3$-$1$~K, $T_K\sim 50$-$200$~mK, and $T\sim 10$-$40$~mK. The gap $\Delta_{\rm topo}$ is generally smaller than the parent gap ($\approx 1.2$~meV for Nb ~\cite{Uday2024}) and must be determined independently. Because $\delta_{\rm edge}$ and $E_C$ both decrease with island size, the window $\delta_{\rm edge}\ll k_BT\ll E_C$, set mainly by the edge velocity and dielectric environment, is likely the most demanding requirement. Finally, once CMCK criticality is observed, tuning the magnetization back to the QAH phase should recover conventional CK in the same device, allowing the universal fingerprints in Table~I, including the residual entropy from charge measurements~\cite{HanC2022,Piquard2026}, to be compared within one sample.

\emph{Summary.} We have shown that a charged island hosting a single chiral Majorana edge mode realizes a distinct quantum critical point, the CMCK critical point. Majorana self-conjugacy locks normal and anomalous tunneling, effectively doubling the Kondo screening channels without increasing the number of physical leads. Consequently, a two-terminal device exhibits four-channel Kondo criticality, with universal exponents and residual entropy distinct from conventional two-terminal CK. For three or more leads, even the critical conductance distinguishes CMCK from conventional CK with the same number of leads. Tuning the same island to its QAH phase provides an in situ conventional-CK baseline, allowing these universal many-body signatures to serve as a potential test of chiral Majorana edge modes using established charge-Kondo techniques.

\emph{Acknowledgment.} I thank Prof. Patrick Lee for useful comments, and acknowledge support from the Quantum Materials and Sensing Institute and from Prof. Arun Bansil. This work was initiated at Oak Ridge National Laboratory with support from the U.S. Department of Energy, Office of Science, National Quantum Information Science Research Centers, Quantum Science Center.

\bibliography{ref}

\end{document}